\documentclass[12pt,preprint]{aastex}
\usepackage{epsf}
\usepackage{natbib}

\tabletypesize{\scriptsize}

\newcommand{\etal}{{et al.}\ }

\def\gappeq{\mathrel{ \rlap{\raise.5ex\hbox{$>$}}
                      {\lower.5ex\hbox{$\sim$}}  } }

\begin{document}
\shorttitle{An Algorithm for Modeling Collisional Dusty Debris Disks}
\shortauthors{Stark \etal}

\title{A New Algorithm for Self-Consistent 3-D Modeling of Collisions in Dusty Debris Disks}

\author{Christopher C. Stark\altaffilmark{1} and Marc J. Kuchner\altaffilmark{2}}

\altaffiltext{1}{Department of Physics, University of Maryland, Box 197, 082 Regents Drive,
College Park, MD 20742-4111, USA;
starkc@umd.edu}
\altaffiltext{2}{NASA Goddard Space Flight Center, Exoplanets and Stellar
Astrophysics Laboratory, Code 667, Greenbelt, MD 20771}

\begin{abstract}

We present a new ``collisional grooming" algorithm that enables us to model images of debris disks where the collision time is less than the Poynting Robertson time for the dominant grain size.  Our algorithm uses the output of a collisionless disk simulation to iteratively solve the mass flux equation for the density distribution of a collisional disk containing planets in 3 dimensions.  The algorithm can be run on a single processor in $\sim$1 hour.  Our preliminary models of disks with resonant ring structures caused by terrestrial mass planets show that the collision rate for background particles in a ring structure is enhanced by a factor of a few compared to the rest of the disk, and that dust grains in or near resonance have even higher collision rates.  We show how collisions can alter the morphology of a resonant ring structure by reducing the sharpness of a resonant ring's inner edge and by smearing out azimuthal structure.  We implement a simple prescription for particle fragmentation and show how Poynting-Robertson drag and fragmentation sort particles by size, producing smaller dust grains at smaller circumstellar distances.  This mechanism could cause a disk to look different at different wavelengths, and may explain the warm component of dust interior to Fomalhaut's outer dust ring seen in the resolved 24 $\mu\rm{m}$ \emph{Spitzer} image of this system.

\end{abstract}

\keywords{circumstellar matter --- interplanetary medium --- methods: N-body simulations --- methods: numerical --- planetary systems}

\section{Introduction}

Recent resolved images of several debris disks reveal complex structures in the form of rings, gaps, and warps carved in the circumstellar dust \citep[e.g.][]{g98,w02,kgc05,g06,s09}.  Some of these structures are likely the result of planetary companions that gravitationally perturb the disk \citep[e.g.][]{q06,c09}.  Modeling these structures can potentially reveal the physical and orbital parameters of the planets, dust grains, and sources of dust in these systems, helping us better understand the late stages of planet formation that debris disks represent \citep[e.g.][]{z01,w02, mkh04, dm05, w07, sk08}.

So far all resolved debris disks are collisionally-dominated systems, meaning the collision time, $t_{\rm coll}$, is much shorter than the Poynting-Robertson (PR) time, $t_{\rm PR}$, in these disks \citep{w05}.  For example, $t_{\rm PR} / t_{\rm coll}$ is $\sim$300 for 10 $\mu\rm{m}$ grains in the resolved circumstellar dust rings of Fomalhaut \citep{c09} and HR 4796A \citep{dws08,s09}.  Many different modeling techniques have been applied to these disks.  However, no model has yet been able to accurately treat gravitational dynamics and collisions simultaneously in a self-consistent fashion.

Some models of dust in debris disks ignore collisions altogether \citep[e.g.][]{mkh04,dm05}.  Some variations on collisionless disk models simply stop the integration of the particle orbits once dust grains have lived as long as their collisional time \citep[e.g.][]{c09}.  The collision time is typically estimated as $t_{\rm coll} = t_{\rm orbit} / \alpha\tau$, where $t_{\rm orbit}$ is the orbital period, $\tau$ is the optical depth of the disk, and $\alpha$ is a constant determined by the distribution of the grains' inclinations and eccentricities and the size threshold for catastrophic collisions, estimated by \citet{wdt99} to be $\alpha \approx 4\pi$.  These models do not include the influence of disk asymmetries and orbital resonances on collision rates.

Other models of dusty disks avoid treating the resonant dynamics of the grains altogether, opting instead to investigate the long-term collisional evolution of debris disks with simple geometries.  Analytic models of azimuthally symmetric steady-state collisional disks with a single belt of parent bodies help us develop our intuition about these systems \citep[e.g.][]{w99,w07}.  Numerical kinetic models, which treat collisions with great detail by including processes such as grain fragmentation and cratering \citep[e.g.][]{tab03, lkr08}, specialize in the long-term behavior of the radial and grain size distributions of dust in disks, and do not include planets.  The hybrid model of \citet{bk06} merges kinetic models with $n$-body models by combining a multi-annulus kinetic model which treats small dust grains with an $n$-body model to treat large planetesimals.  However, this code is tailored for grain growth simulations and not resonant structures in debris disks; it does not model the resonant dynamics of the dust grain population.

Still other models look at only the transient effects of collisions.  The collisional code of \citet{gat07} is designed for modeling collisional avalanche events in debris disks, but does not have the spatial resolution necessary to model structures caused by gravitational resonant dynamics.  Robust many-particle simulations that model individual collisional events and follow the orbits of any fragments produced are currently limited by computer processing power to $\sim$$10^4$ particles for long-term integrations \citep[e.g.][]{l09}, restricting their debris disk application to short-term integrations or integrations sampling limited phase space.

Here we present a novel ``collisional grooming" algorithm for treating collisions and resonant dynamics self-consistently in an optically thin disk.  The algorithm produces a dust distribution for each grain size that simultaneously solves the equation of motion for small dust grains,
\begin{equation}
	\label{equationofmotion}
  	\frac{d^2{\bf r}}{dt^2} = -\frac{Gm_{\rm \star}}{r^2}(1-\beta)\hat{\bf r}-\frac{(1+{\rm sw})\beta}{c}\frac{Gm_{\rm \star}}{r^2}[{\dot{r}}{\hat{\bf r}} + {\bf v}] +\displaystyle\sum_i \frac{Gm_i}{|{\bf r}_i - {\bf r}|^3}({\bf r}_i - {\bf r})\; ,
\end{equation}
and the particle number flux equation in 3D,
\begin{equation}
	\label{numberfluxequation}
	\nabla \cdot \left( n \overline{{\bf v}}\right) - \frac{\partial n}{\partial t} = \frac{\partial n}{\partial t} \bigg|_{\rm coll}\; .
\end{equation}
\noindent Here $G$ is the gravitational constant, $m_{\rm \star}$ is the stellar mass, $c$ is the speed of light, $\textbf{r}$ and $\textbf{v}$ are the heliocentric position and velocity of the grain, $\beta$ is the ratio of radiation pressure force to gravitational force on a grain, ${\rm sw}$ is the ratio of solar wind drag to PR drag, and $m_i$ and ${\bf r}_i$ are the mass and heliocentric position of the $i^{\rm th}$ planet.  The particle number density, $n$, the mean flow velocity, $\overline{{\bf v}}$, and the particle number density removed or added by collisions per unit time, $\partial n/\partial t |_{\rm coll}$, may be functions of position and grain size.

Our algorithm uses a collisionless disk simulation (a ``seed model") as input to calculate initial collision rates, similar to the method developed by \citet{cm03}.  However, our algorithm differs from that of \citet{cm03} in several ways.  Our algorithm includes Poynting-Robertson drag and is therefore applicable to systems with small dust grains.  Our algorithm also allows for the possibility that collisions can affect the dynamics of the system, and it uses an iterative scheme to find the correct density distribution for a steady-state disk.  After the integration of the seed model, the algorithm can be run on a single processor in $\sim$1 hour.  This algorithm can generate new models that should allow us to interpret images of collision-dominated disks like those orbiting Fomalhaut, Vega, Epsilon Eridani, and HR 4796A quantitatively for the first time.


\section{Numerical Method}

\subsection{Collisionless seed models}

We first run a seed model, a model of a steady-state collisionless debris disk.  We numerically integrate Equation \ref{equationofmotion} for a collection of particles launched from parent bodies orbiting a star.  This equation includes the dynamical effects of gravity, radiation pressure, corpuscular drag, and PR drag \citep[see][for details]{sk08}.  The drag forces cause the particles to slowly lose angular momentum and spiral inward from the parent bodies from which they were launched.

During their journey inward toward the star, the particles can become temporarily trapped in the external mean motion resonances (MMRs) of any planets present, creating an overdense circumstellar ring structure near the planet's orbit.  We use enough particles, typically on the order of a few thousand, to accurately populate the external MMRs of the planets \citep{sk08}.  During the integration, we record the barycentric coordinates of each particle at a regular interval, $t_{\rm record}$.  This commonly-used technique extrapolates the results of a few thousand particles to millions of particles \citep[e.g.][]{d94, lz99, mmm02, mkh04}.  Our algorithm requires the local velocity distribution to calculate the local collision rate, so we also record the barycentric velocities of each particle at the same interval, $t_{\rm record}$.

\begin{figure}
\begin{center}
\includegraphics[height=6.5in,angle=270]{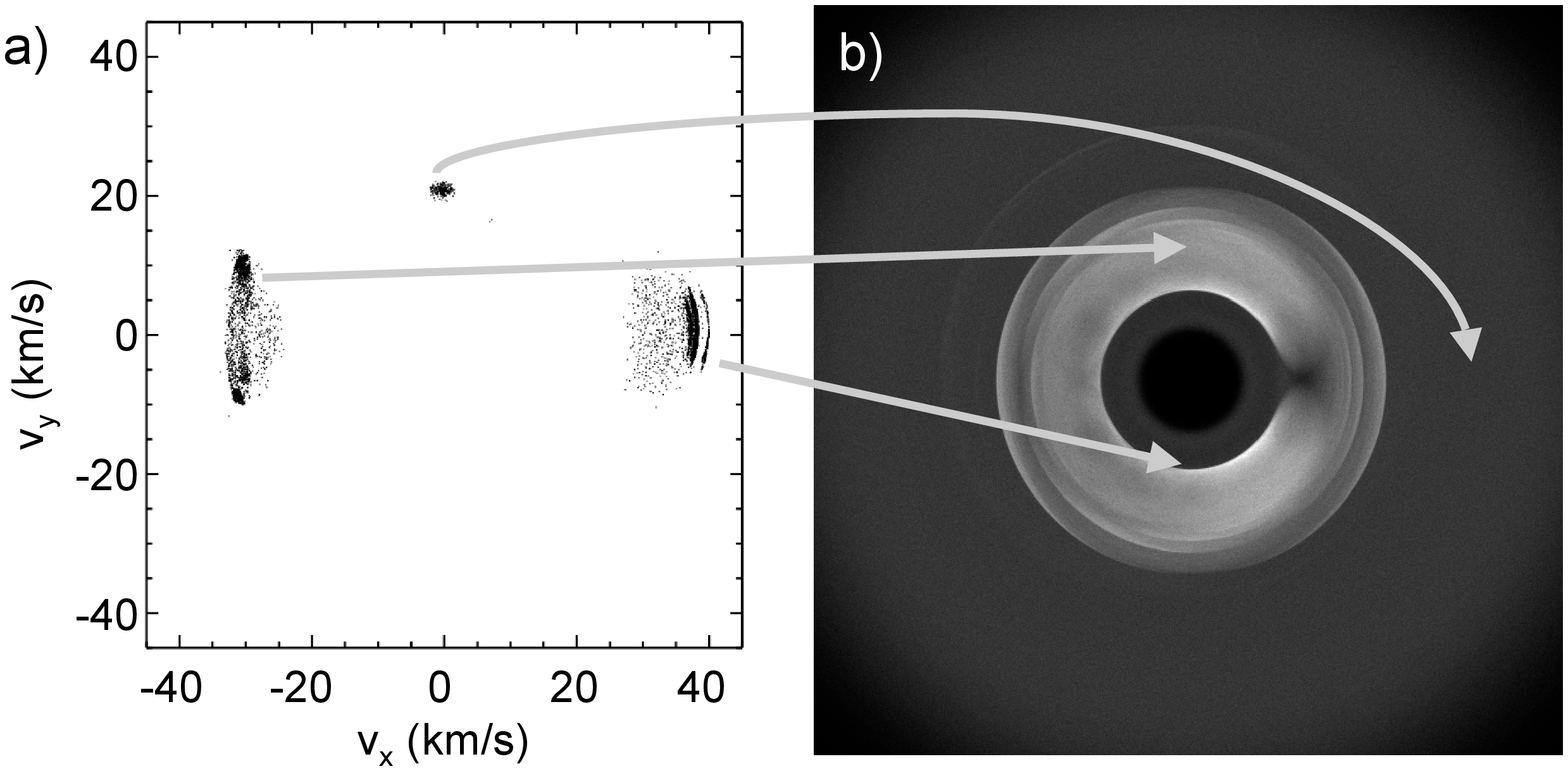}
\caption{Velocity distributions (\emph{a}) at three locations (\emph{b}) in a collisionless dust disk with a resonant ring structure caused by an Earth-mass planet at 1 AU.  The velocity distributions in the resonant ring structure vary greatly and are non-Gaussian. \label{velocity_distribution_figure}}
\end{center}
\end{figure}

We then place all of the records of the barycentric coordinates and velocities into a 3D spatial grid of bins, forming a 3D histogram that represents the distribution function for the collisionless system.  Panel \emph{b} in Figure \ref{velocity_distribution_figure} shows the histogram of the particle density viewed face-on for a collisionless disk model with a resonant ring created by an Earth-mass planet on a circular orbit at 1 AU around the Sun, using a grid bin size of 0.05 AU.  Panel \emph{a} shows the velocity records from three of these bins located in the midplane of the disk, pointed to in panel \emph{b}.

Exterior to the resonant ring structure in Figure \ref{velocity_distribution_figure}, the velocity distribution is approximately Gaussian and the dispersion in the radial direction is approximately twice the dispersion in the azimuthal direction, as expected for a Keplerian disk \citep[e.g.][]{bt87}.  However, within the resonant ring structure the velocity distribution is highly non-Gaussian and the velocity dispersion varies greatly from one location to another.  This illustration shows that to calculate the collision rates in a resonant ring we must explicitly calculate the local velocity distribution.

\subsection{The collisional grooming algorithm \label{implementation}}

Besides the distribution function, recording the barycentric coordinates and velocities of all particles at regular intervals during the integration of the seed model yields a second important ingredient for our algorithm---a chronological record of each particle's trajectory, which we refer to as a ``stream."  We let each stream from our seed model represent a large number of particles, which varies from record to record, i.e. the $i^{\rm th}$ record is scaled to $N_i$ particles.  We adjust the scaling factor to control the mean collision time of the particles.

We initially assume collisions only serve to remove material from a stream.  As they progress through the cloud, the streams become attenuated by collisions with other streams as
\begin{equation}
	\label{attenuationequation}
	N_{i} = N_{i-1}\, e^{-\tau_{{\rm coll,}i}},
\end{equation}
where $N_i$ is the number of particles in the $i^{\rm th}$ record of a given stream and $\tau_{{\rm coll,}i}$ is the collision depth for the $i^{\rm th}$ record.  Equation \ref{attenuationequation} is analogous to the solution to the radiative transfer equation for photons passing through an absorptive medium.  We approximate the collisional depth as 
\begin{equation}
	\label{collisiondepthequation}
	\tau_{{\rm coll,}i} \approx \displaystyle\sum_k n_k \sigma_k |{\bf v}_{i} - {\bf v}_k| t_{\rm record},
\end{equation}
where ${\bf v}_{i}$ is the velocity associated with the $i^{\rm th}$ record of the given stream, $n_k$, $\sigma_k$, and ${\bf v}_k$ are the particle number density, collisional cross-section, and velocity of the other records in the same bin as the $i^{\rm th}$ record, and $n_k$ is  equal to $N_k$ divided by the bin volume.  This approximation works as long as $t_{\rm record} \ll t_{\rm coll}$.  We perform this calculation for all records in all particle streams, one at a time, from the first record to the last record in each stream.

After only one pass through all of the records for all streams, i.e., one iteration, the particle streams will be attenuated incorrectly; the streams will be attenuated based on the density distribution of the collisionless seed model, which overestimates the particle density and therefore the collision rate.  To remedy this problem, we iterate the attenuation process of all records until no record changes by more than a set tolerance of a few percent.  By doing so, we ensure that the final number density histogram approximately satisfies Equation \ref{numberfluxequation} in the steady state.

\begin{figure}
\begin{center}
\includegraphics[width=6.5in]{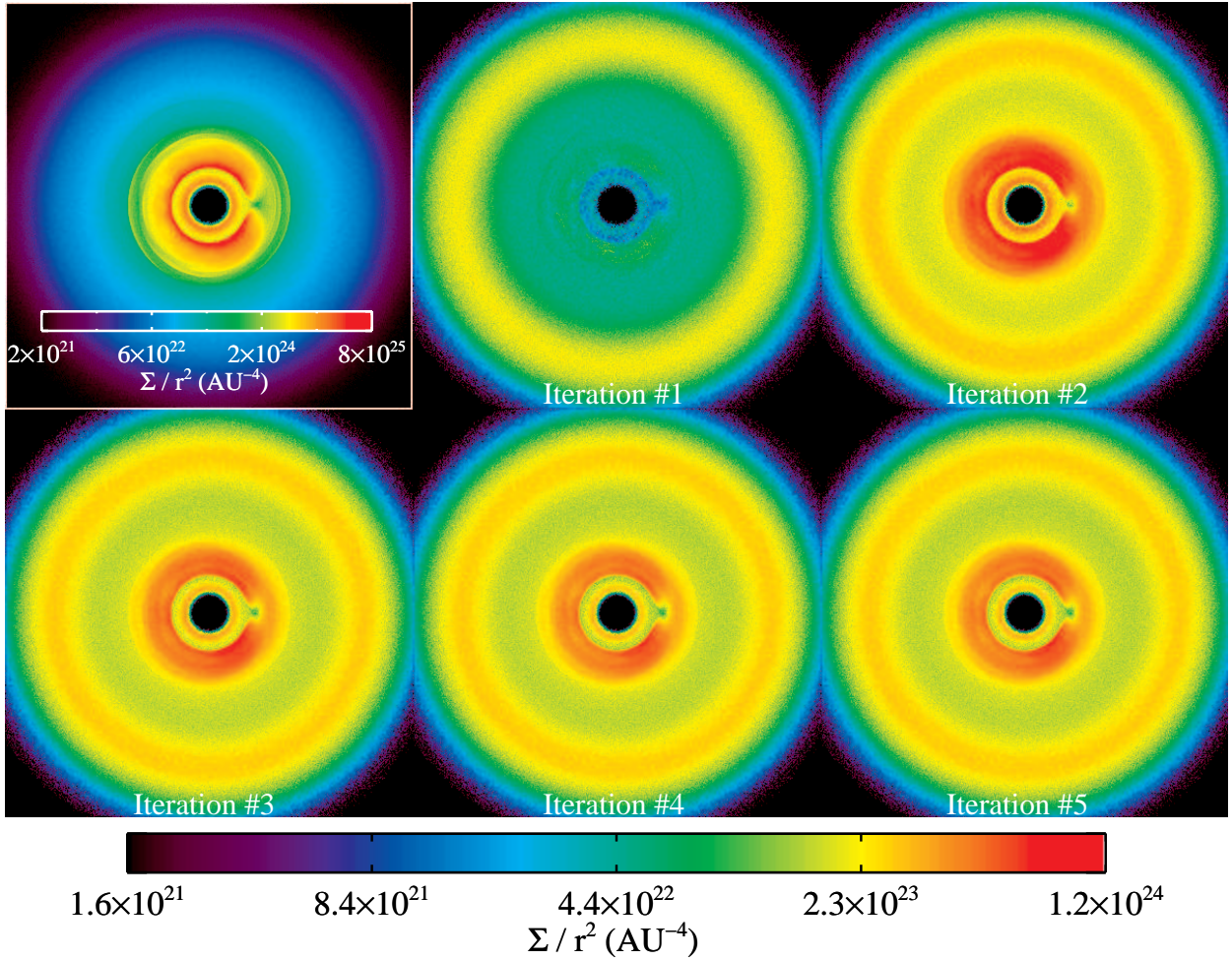}
\caption{Surface density scaled by $1/r^2$ of a collisional model as a function of iteration number.  The upper-left panel shows the surface density of the seed model, a collisionless disk with a resonant ring structure (see Section \ref{test_section} for simulation details).  The iterations alternate between over- and underestimating the amount of collisions while converging on the correct solution. \label{density_vs_iteration_figure}}
\end{center}
\end{figure}

Figure \ref{density_vs_iteration_figure} shows an example of the grooming algorithm at work.  The top-left panel shows the surface density of a collisionless disk scaled by $1/r^2$, where $r$ is circumstellar distance.  The disk of 120 $\mu\rm{m}$ grains features a ring structure caused by an Earth-mass planet on a circular orbit at 1 AU around a Sun-like star.  We applied our collisional grooming algorithm to the disk and scaled the number of particles per stream equally among all streams such that $\eta_0 \approx 3.7$, where $\eta_0 = t_{\rm PR}(r_0) / t_{\rm coll}(r_0)$ and $r_0$ is the mean circumstellar distance at which grains are launched.

After the first iteration, shown in the top-middle panel, the collision rates are overestimated so that the surface density in the inner disk is too low.  During the second iteration, shown in the top-right, the algorithm underestimates the collision rates.  The algorithm alternates between over- and underestimating the collision rates while converging on the correct solution.

Our algorithm uses a finite grid of bins to approximate the local density and velocity structure, so Poisson noise can become an issue in bins with few records.  To help mitigate this Poisson noise, we use a nearest-neighbor averaging routine.  For bins with fewer than 10 records, we include the records from the six nearest-neighbor bins in the calculation of $\tau_{\rm coll}$.  We weight the records in each nearest-neighbor bin at 50\% of the weight of the central bin.

Our calculation of the collisional depth (Equation \ref{collisiondepthequation}) trivially handles seed models with multiple grain sizes.  Given an initial distribution of grain sizes, the algorithm can simultaneously solve for the collisional interactions among all grains of all sizes included in the seed model.  The algorithm can model complex phenomena that depend on grain size, such as size-dependent collision rates, radial transport rates, and resonant structure morphologies.

Equation \ref{attenuationequation} ignores any fragments that may be produced by collisions.  Treating collisions as non-productive is not valid for all collisions in all debris disks, but it is likely acceptable in a wide range of cases.  \citet{bp93} argued that in the inner solar system, grain-grain collision velocities are high enough (on the order of a few kilometers per second) that grains fragment catastrophically during any collision, and any fragmentation products are either gas or small enough to be removed immediately by radiation pressure.  In resonant ring structures, like the circumsolar ring created by Earth, the collision velocities can be even higher, on the order of tens of kilometers per second (c.f. Figure \ref{velocity_distribution_figure}).  In the outer solar system, where collision velocities are lower, grains probably resemble cometary grains.  Samples of cometary interplanetary dust particles directly returned from the Stardust mission and observations of cometary ejecta during the Deep Impact mission reveal that the majority of observed cometary particles are loosely bound aggregates of submicron-sized grains, which can easily be shattered into unbound $\beta$-meteoroids \citep{ah05, b06, z06}.

Our algorithm can also be adapted to handle particle fragmentation.  We discuss this feature in Section \ref{particlefragmentationsection}.

\subsection{Tests \label{test_section}}

We developed a collisional disk modeling code based upon the algorithm described above.  We subjected our code to a battery of tests to confirm its operation and identify its limits.  Here we demonstrate the algorithm's performance, show that it converges on a unique and correct solution, and place limits on the conditions under which it is valid.  For now, we neglect fragmentation; we assume non-productive collisions.

Below, we will refer to an analytic solution for the surface density as a function of circumstellar distance for a planet-less disk.  \citet{w99} showed that Equation \ref{numberfluxequation} has the following steady-state solution for an azimuthally-symmetric disk with a single grain size and grain mass under the assumption that collisions create no daughter particles:
\begin{equation}
	\label{analyticsolutionequation}
	\Sigma(r) = \frac{\Sigma(r_0)}{1 + 4 \eta_0 (1 - \sqrt{r/r_0})},
\end{equation}
where $\Sigma$ is the surface density and $r_0$ is the circumstellar distance of the dust source.  If we do not add a perturbing planet, we can directly compare the results of our algorithm to this expression.

\subsubsection{Seed models \label{seedmodelsection}}

Throughout this paper, we will refer to two collisionless disk simulations which we use as seed models: a planet-less disk simulation and a simulation of a disk with a resonant ring structure.  For both simulations, we integrated the orbits of 20,000 particles.  Particle integrations were terminated when their semi-major axes were less than 0.5 AU, so many images will show no data interior to this radius.  In many instances throughout this paper we will only examine a subset of the total number of integrated particles, and will state the number used when appropriate.  All simulations were performed using a hybrid symplectic integrator \citep[see][]{sk08}.

Our planet-less disk seed model contains particles with $\beta = 0.0023$, where $\beta$ is the ratio of the force on the grain from radiation pressure to the gravitational force \citep[e.g.][]{bls79}.  For a spherical blackbody grain with a density of 2 gm cm$^{-3}$, $\beta = 0.0023$ corresponds to a grain radius of 120 $\mu{\rm m}$ around a Sun-like star.  We initially placed the grains at 10 AU on circular orbits with inclinations uniformly distributed between 0 and 14$^{\circ}$.  We distributed all other initial orbital parameters (longitude of ascending node, mean anomaly, argument of pericenter) uniformly between 0 and 2$\pi$.  We recorded particle positions and velocities every 6956 years.

For the second seed model, we integrated the orbits of particles with $\beta = 0.0023$ as they spiraled inward and interacted with an Earth-mass planet on a circular orbit at 1 AU around a Sun-like star.  We launched grains from parent bodies with semi-major axes uniformly distributed between 3.5 and 4.5 AU, eccentricities uniformly distributed between 0 and 0.2, and inclinations uniformly distributed between 0 and 20$^{\circ}$.  We distributed all other initial orbital parameters uniformly between 0 and 2$\pi$.  We recorded the particle positions and velocities once every 426 years.

\subsubsection{Bin size test}

The finite size of the bins used to approximate the local density and velocity distributions is a natural source of error for our algorithm.  We need to ensure that the bins are small enough to resolve any structure within the disk.  To help us decide on the appropriate bin size, we performed the following test.

We applied our collisional grooming algorithm to our seed model of a disk with a resonant ring structure.  We used 15,000 simulated particles to ensure that Poisson noise was not an issue for this test.  For the collisional algorithm, we scaled the number of particles per stream such that $\eta_0 \sim 1$, and used four different cubic bin sizes of 0.02, 0.05, 0.1, and 0.2 AU.  For each of the largest three cases, we calculated the differences in the collision rates compared to the smallest case.

\begin{figure}
\begin{center}
\includegraphics[width=6.5in]{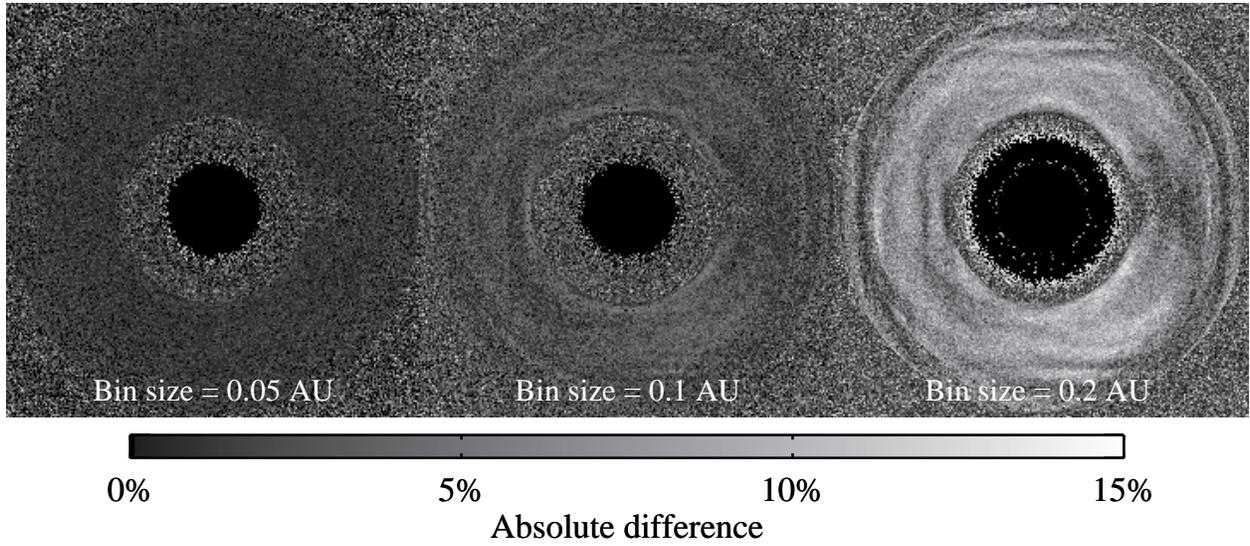}
\caption{Errors in the collision rate in a model with an Earth-mass planet at 1 AU.  Errors are relative to the collision rates calculated using a bin size of 0.02 AU. \label{binsize_test_figure}}
\end{center}
\end{figure}

Figure \ref{binsize_test_figure} shows these relative differences in the collision rates.  The left panel shows that differences in the collision rates between the 0.05 and 0.02 AU bin sizes are on the order of a few percent or less, and show no signs of structural differences.  The right panel shows that using a bin size of 0.2 AU results in obvious structural differences (greater than 10\%) compared to a model with a 0.02 AU bin size.  The right panel also shows a subtle imprint of the grid itself, which appears as straight horizontal and vertical features in the collision rates.

In light of these results, we recommend that the bin size for collisional calculations in a debris disk with a resonant ring structure should be $\sim0.05$ AU for a planet at 1 AU.  The size of structural features in a resonant ring scale linearly with planet semi-major axis, $a_{\rm p}$ \citep{sk08}, so we suggest that in general, the optimal bin size for disks with resonant ring structures is $\sim 0.05\,a_{\rm p}$.  Bins larger than this size fail to resolve the ring structure while bins smaller than this size become more susceptible to Poisson noise in the distribution function.

For collision times that are short compared to the PR time ($\eta_0 \gg 1$), grains launched from circumstellar distance $r_0$ have little time to move radially inward before colliding.  This process can result in a very narrow ring structure at $r=r_0$.  If the bin size is larger than the width of the ring, our collisional algorithm will fail to resolve the ring.  So we must also choose a bin size small enough such that the crossing time of a bin caused by PR drag at $r=r_0$ is shorter than the collision time.  The bin crossing time is given by
\begin{equation}
	t_{\rm cross}(r_0) = t_{\rm PR}(r_0) - t_{\rm PR}(r_0-b)\; ,
\end{equation}
where $b$ is the bin size and the PR time is given by
\begin{equation}
	t_{\rm PR}(r) = \frac{c\,r^2}{4GM_{\star}\beta}\; ,
\end{equation}
where $c$ is the speed of light, $G$ is the gravitational constant, and $M_{\star}$ is the stellar mass \citep{ww50}.  With the requirement $t_{\rm cross}(r_0) \lesssim t_{\rm coll}(r_0)$ we find to first order in $\eta_0^{-1}$ that the bin size must satisfy
\begin{equation}
	\label{binsizeequation}
	b \lesssim \frac{r_0}{2\eta_0}\; .
\end{equation}

\subsubsection{Poisson noise test \label{poisson_noise_section}}

If there are too few records in a given bin, Poisson noise can dominate the calculation of the collisional depth, even with our nearest-neighbor averaging routine.  However, we are not specifically concerned with whether a single collisional depth calculation suffers from Poisson noise, but whether Poisson noise affects the final outcome of the simulation.  To examine the effects of Poisson noise, we applied our collisional algorithm to our planet-less seed model three times, first using 15,000 simulated particles, then using 5,000 simulated particles, and then using only 1,500 simulated particles.  For each application of the collisional algorithm, we used a cubic bin size of 0.05 AU.  We scaled the disk such that $\eta_0 \approx 223$ for all three cases.

\begin{figure}
\begin{center}
\includegraphics[width=6.5in]{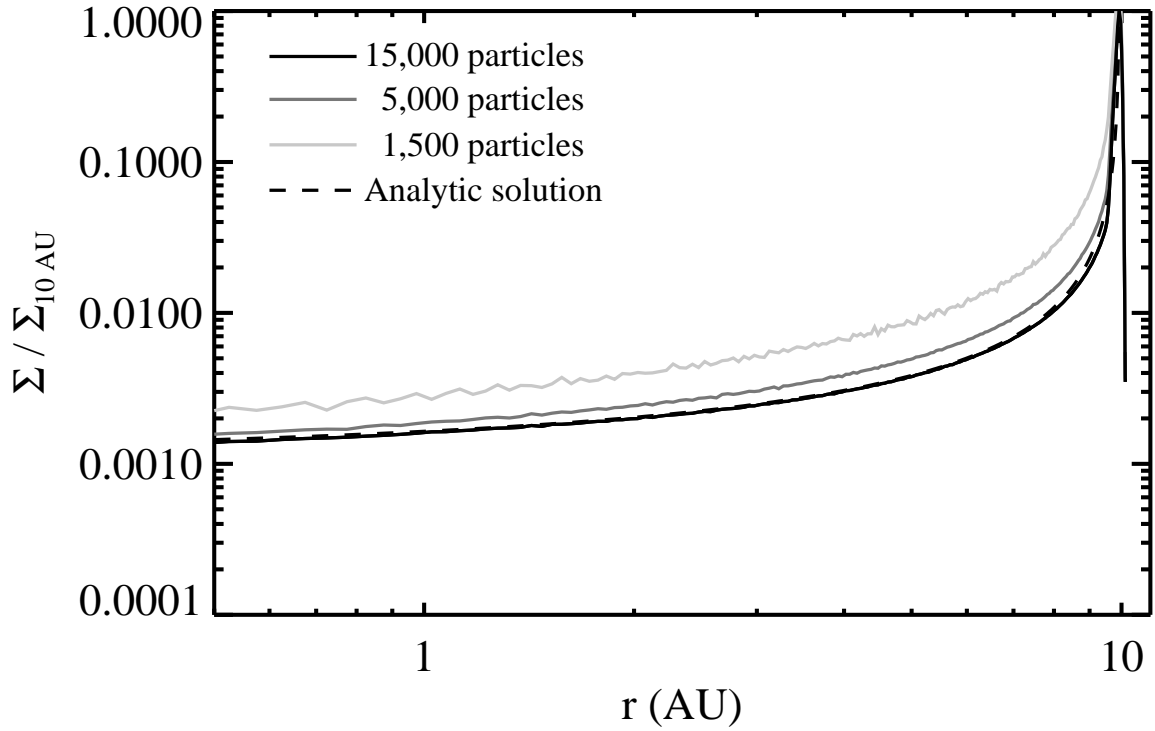}
\caption{Azimuthally averaged surface density as a function of circumstellar distance in a planet-less collisional disk for three simulations with different numbers of particles (see Section \ref{test_section}).  The 15,000-particle simulation follows the analytic solution well, but the 1,500-particle simulation deviates significantly and shows signs of Poisson noise. \label{poisson_noise_test_figure}}
\end{center}
\end{figure}

After processing the three disks with our collisional grooming algorithm, we azimuthally averaged each of the three resulting surface densities.  Figure \ref{poisson_noise_test_figure} shows the normalized surface density as a function of circumstellar distance for all three cases compared to the analytic solution (Equation \ref{analyticsolutionequation}).  The 15,000-particle case follows the analytic solution well, while the 1,500-particle case deviates by a factor of $\sim 2$ and contains strong fluctuations caused by Poisson noise.  From these simulations we estimate that the average number of records per bin near the circumstellar distance at which grains are launched should be at least on the order of a few to avoid the effects of Poisson noise.

\subsubsection{Uniqueness of solution}

Figure \ref{density_vs_iteration_figure} shows that our algorithm does indeed converge, but does not indicate whether the algorithm converges on a unique solution.  To test for uniqueness, we applied the collisional grooming algorithm to three independent seed simulations of disks with ring structures, using 5,000 particles for each simulation and a value of $\eta_0 \sim 4$.  The range of initial conditions were the same for all three simulations, but individual values were generated using a different random number seed.

We compared the resulting surface densities of all three collisional disks.  Except for the outer and inner extremities of the disk where the statistics were poor, none of the final disk surface densities differ by more than a few percent and none show significant structural differences.  All three independent simulations converged to the same surface density solution to within the limits of Poisson noise.

\subsubsection{Correctness of solution \label{analytic_comparison_section}}

\begin{figure}
\begin{center}
\includegraphics[width=6.5in]{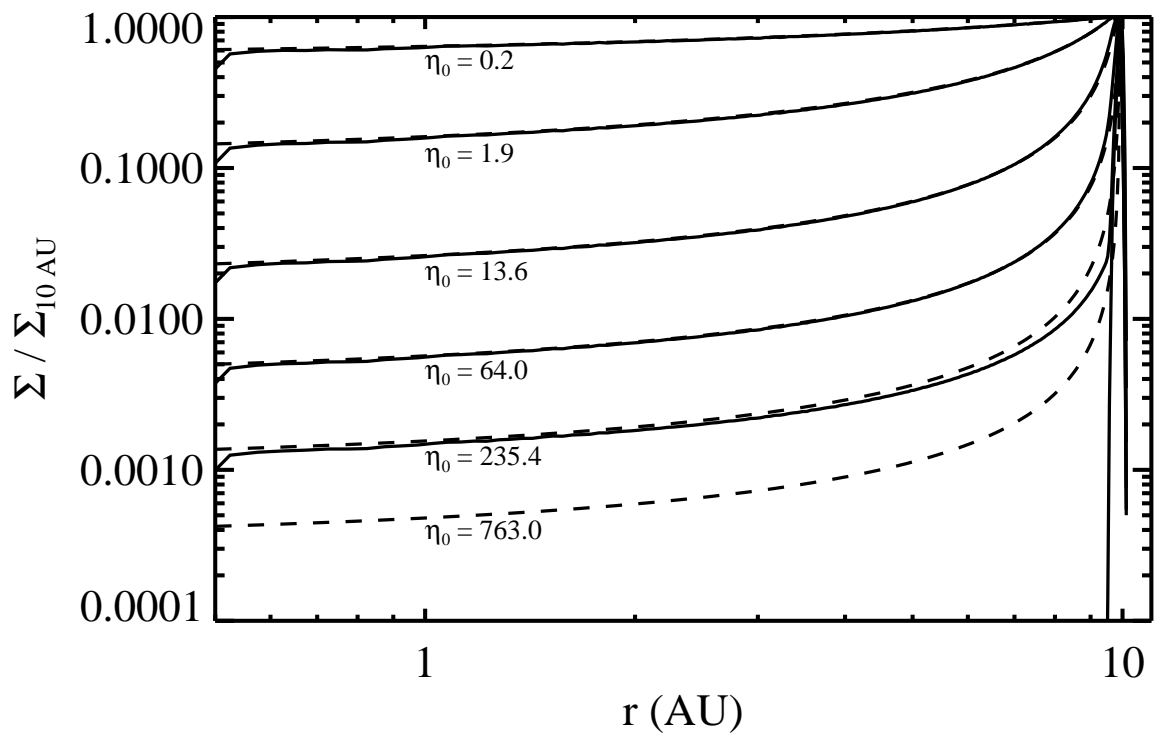}
\caption{Azimuthally averaged steady-state surface density as a function of circumstellar distance for a planet-less disk undergoing PR drag and non-productive collisions.  Analytic solutions are shown with dashed lines and the results of our collisional algorithm are shown with solid lines.  Our algorithm gives the correct solution for all values of $\eta_0$ except the largest, at which point the collisions yield a narrow ring near 10 AU that is no longer resolved by the grid. \label{analytic_comparison_test_figure}}
\end{center}
\end{figure}

To test the correctness of our algorithm's solution, we directly compared the results of our algorithm to the analytic solution for a planet-less disk of a single grain size (Equation \ref{analyticsolutionequation}).  We applied our collisional grooming algorithm to our planet-less-disk seed model of 20,000 particles using six different values of $\eta_0$.  Figure \ref{analytic_comparison_test_figure} shows the azimuthally-averaged surface densities as a function of circumstellar distance for all six cases.  The calculated surface densities (solid lines) match the analytic solutions (dashed lines) well for all values of $\eta_0$ except $\eta_0 = 763$.  In this case, collisions happen so quickly that the disk forms a narrow ring at 10 AU whose width is smaller than the bin size used in our algorithm (0.05 AU); this case does not meet the bin size criteria in Equation \ref{binsizeequation}.  If we used a smaller bin size, we could resolve the ring structure and investigate even larger values of $\eta_0$.

For all cases shown in Figure \ref{analytic_comparison_test_figure}, $t_{\rm record} \ll t_{\rm coll}$; none of the deviations from the analytic values are caused by insufficient time sampling.  We tested the algorithm's behavior when $t_{\rm record} \gtrsim t_{\rm coll}$, and confirmed that our algorithm fails under these circumstances.  For $t_{\rm record} \gtrsim t_{\rm coll}$, our algorithm typically overestimates the collision rate and the result looks qualitatively similar to the $\eta_0 = 763$ case shown in Figure \ref{analytic_comparison_test_figure}.

For the case of a disk with a resonant ring structure, there exists no analytic solution for the surface density with which we can compare the results of our simulations.  However, we can probably assume that our collisional algorithm arrives at the correct solution if the amount of collisions is very small, i.e., $\eta_0 \ll 1$, since such a disk should deviate from the collisionless case by very little.  Under this assumption, we propose the following test to investigate the correctness of our algorithm's solution for a disk with a ring structure:
\begin{enumerate}
	\item Apply our algorithm to the collisionless seed model using $\eta_0 \ll 1$.
	\item Store the output of the algorithm, call it Model C.
	\item Increase $\eta_0$ by a small amount $\delta\eta_0$.
	\item Apply our algorithm to Model C disk using the new value of $\eta_0$.
	\item Repeat steps 2 through 4 until $\eta_0$ is equal to the desired value of $\eta_0$.
	\item Compare the results to a model with the same value of $\eta_0$ calculated in only one step.
\end{enumerate}

We performed this test using a 5,000-particle seed model of a disk with a ring structure.  We used a cubic bin size of 0.05 AU and scaled the disk density so that the final $\eta_0 \approx 3.7$.  We applied the collisional grooming algorithm using 19 logarithmically-spaced steps in $\eta_0$.  We compared the surface density of this disk to the surface density calculated by applying the algorithm in the usual single-step fashion.  The disks differed by less than one part in $\sim 10^5$; both methods arrived at the same solution to within the limits of Poisson noise.  A similar test performed in the opposite direction, i.e., slowly reducing the $\eta_0$ to the desired value, gave similar results.  This test supports both the uniqueness and correctness of the solution that our algorithm finds for a collisional disk with a resonant ring structure.

\subsubsection{Benchmark tests}

We benchmarked our collisional grooming algorithm code by applying it to our seed model of a disk with a ring structure.  We recorded the run time of our code using 1,250, 2,500, 3,750, and 5,000 simulated particles and cubic bin sizes of 0.05 and 0.1 AU.  Table \ref{benchmark_table} shows the run time for a single iteration of our algorithm on a single 2.2 GHz CPU.

For each bin, the algorithm performs of order $n_{\rm b}^2$ calculations, where $n_{\rm b}$ is the number of records in that bin.  So we would expect that our run time per iteration scales as $B \langle n_{\rm b}\rangle ^2$, where $B$ is the number of bins containing records, or as $\langle n_{\rm b}\rangle ^2$ for a given bin size.  Table \ref{benchmark_table} shows that the run time scales as $\langle n_{\rm b}\rangle ^2$ for a bin size of 0.1 AU, but not for a bin size of 0.05 AU.  In the latter case, our algorithm is working with many bins that have relatively few entries and is switching on our nearest-neighbor approximation described in Section \ref{implementation}, which can cause the run time to deviate from the $\langle n_{\rm b}\rangle ^2$ scaling relationship.

The number of iterations required to converge on the correct solution depends on many factors, such as $\eta_0$, the number of records, the bin size, the structure of the disk, and the tolerance set for convergence.  For a disk with a resonant ring structure composed of 5,000 particles, typically 5 -- 10 iterations are required for better than 5\% convergence with $\eta_0 \sim 1$.  The total run time for such a disk typically ranges from 20 minutes to 2 hours.

\subsection{Limitations}

Here we summarize the current limitations of our algorithm.  We have already shown that the bin size must be small enough to resolve any structural features in the disk, and the collision time must be longer than the time between records for our algorithm to converge.  We also showed that the number of records per bin must be large enough to avoid the effects of Poisson noise.

The algorithm presently has another, more subtle limitation, which it shares with many collisionless steady-state disk simulations \citep[e.g.][]{d94, lz99, mkh04,dm05}.  As stated in Section \ref{implementation}, we record the coordinates and velocities of each integrated particle at regular time intervals, $t_{\rm record}$, to extrapolate the results of a few thousand integrated grains to millions of grains or more and to obtain a chronological record of each particle's trajectory.  This technique implies that a new parent body launches grains every $t_{\rm record}$ with the exact same orbital parameters as the first parent body.  Because our algorithm requires the use of this technique, our algorithm implicitly assumes that each parent body's orbit is populated by many parent bodies uniformly distributed in mean anomaly and that the orbits of parent bodies do not change over the course of a collisional time.


\section{Non-productive collisions in resonant ring structures}

We can use our algorithm to investigate the effects of collisions on steady-state resonant ring structures.  To this end, we ran some models of familiar kinds of disk structures. Here we discuss some of the new physical effects we have observed in our models.

\begin{figure}
\begin{center}
\includegraphics[width=6.in]{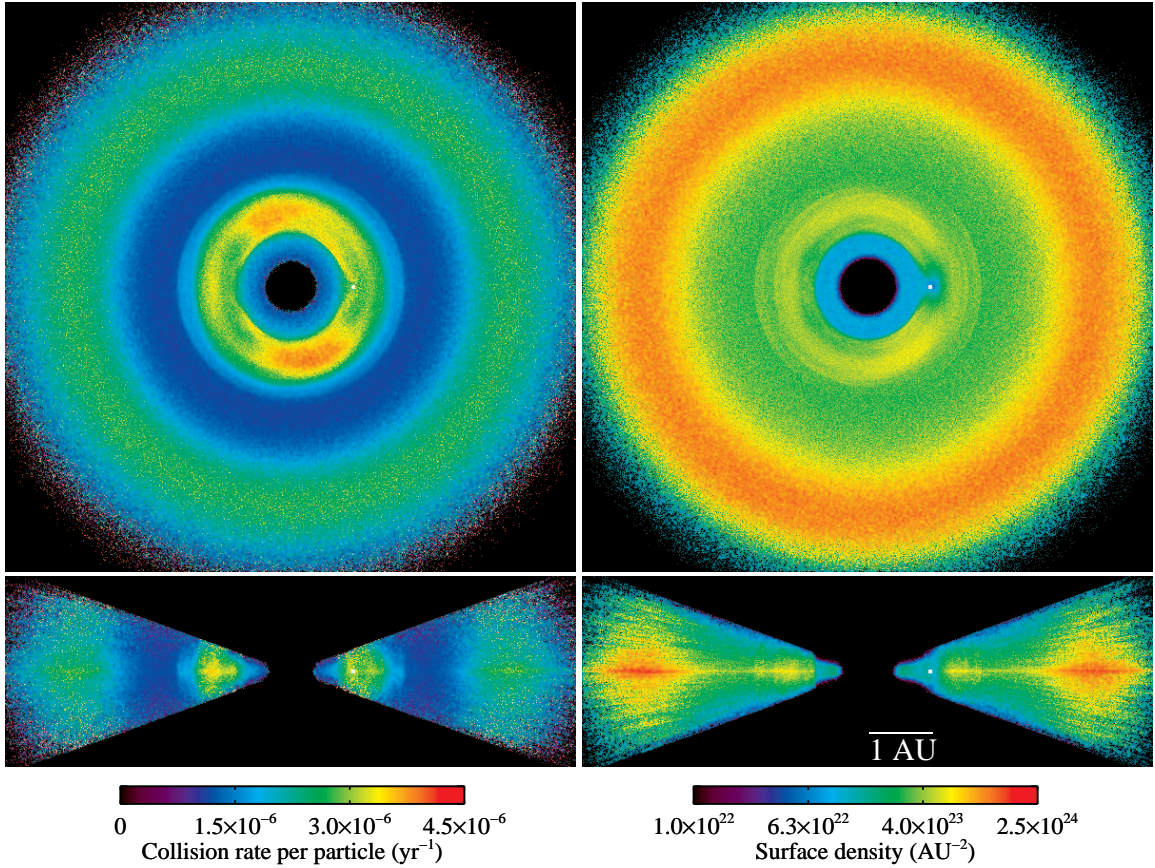}
\caption{\emph{Top-left:} Collision rate per particle ($|dN/Ndt|$) in a 0.4 AU-thick cross-section through the midplane of a disk with a resonant ring structure.  A white dot marks the location of an Earth-mass planet orbiting at 1 AU.  \emph{Top-right:} Surface density of the midplane cross-section shown in the upper-left panel.  \emph{Bottom-left:} Collision rate per particle in a 0.4 AU-thick edge-on cross-section of the same disk.  \emph{Bottom-right:} Surface density of the edge-on cross-section shown in the bottom-left panel.  The collision rate is affected by both the local density structure and the local velocity distribution; the collision rate is highest in the resonant ring structure, a region of enhanced density and relative particle velocities.\label{collision_rate_faceon_figure}}
\end{center}
\end{figure}

The top-left panel of Figure \ref{collision_rate_faceon_figure} shows the collision rate per particle in a 0.4 AU-thick cross-section through the mid-plane of a disk.  The disk has a resonant ring structure caused by an Earth-mass planet whose location is marked with a white dot.  We scaled the disk density so that $\eta_0 \approx 3.7$.  The top-right panel shows the surface density of the same cross-section for comparison.

The collision rate is high in a circumstellar ring near the location of the parent bodies ($\sim$3 AU), i.e., the birth ring.  Grains in this region of the disk are too young to have been destroyed by collisions, so the local density is relatively high, as seen in the top-right panel.  The collision rate drops just interior to this ring, at a circumstellar distance of $\sim$2 AU.

As you might expect, the top-left panel of Figure \ref{collision_rate_faceon_figure} shows that the collision rate reaches its highest point in the resonant ring structure, where the density is enhanced by resonant trapping and relative velocities are higher because of resonant pumping of the grains' eccentricities.  The average collision rate in the resonant ring structure is higher by a factor of a few when compared to the collision rate exterior to the resonant ring structure, consistent with analytic estimates \citep{q07}.  Figure \ref{collision_rate_faceon_figure} also shows that the collision rate exhibits azimuthal and radial structure in the resonant ring.  This structure reflects both localized density enhancements and regions of higher velocity dispersion.  For example, the region of enhanced collision rate located $\sim$90$^{\circ}$ clockwise from the planet generally corresponds to a region of higher density.  But the region of enhanced collision rate located $\sim$90$^{\circ}$ counterclockwise from the planet does not.  This second region of enhanced collision rate is primarily caused by an increase in the local velocity dispersion, as shown in Figure \ref{velocity_distribution_figure}. 

We show the collision rate and surface density for an edge-on cross-section of the same disk in the bottom two panels of Figure \ref{collision_rate_faceon_figure}.  The bottom-left panel reveals a trend toward higher collision rates in the disk mid-plane, which is denser than the rest of the disk, as shown in the bottom-right panel.  The bottom-left panel of Figure \ref{collision_rate_faceon_figure} also shows that the collision rate at a circumstellar distance of $\sim$0.7 AU is higher than the collision rate at $\sim$2 AU by a factor of $\sim$2, even though the density is higher near $\sim$2 AU.  This increase in the collision rate occurs because grains that survive and spiral inward past the resonant ring structure have typically had their eccentricities pumped up by passage through the resonance, so the velocity dispersion interior to the resonant ring structure is higher than the velocity dispersion exterior to the resonant ring structure.

\begin{figure}
\begin{center}
\includegraphics[width=5.5in]{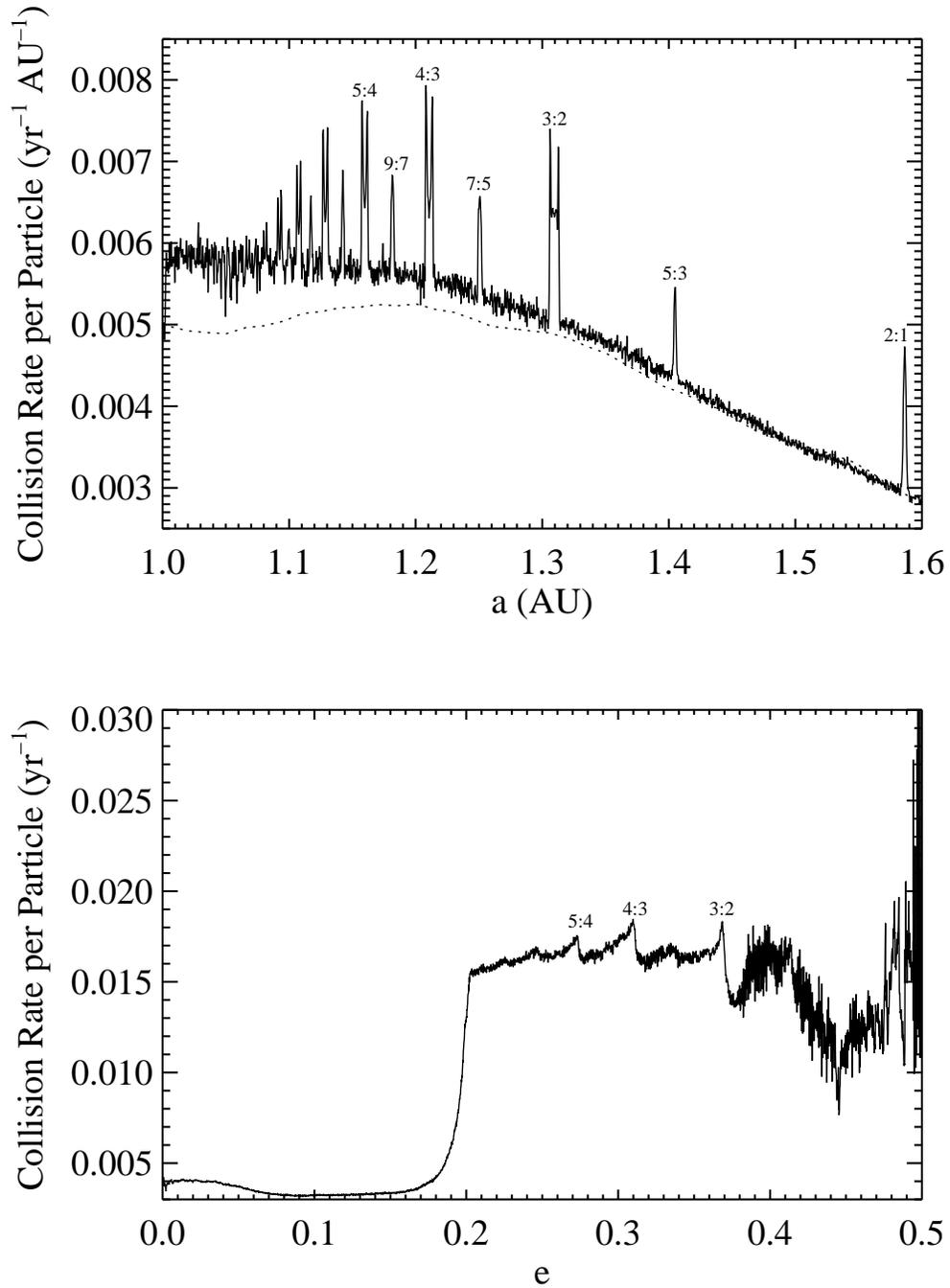}
\caption{Collision rate per particle ($|dN/Ndt|$) as a function of semi-major axis and eccentricity for a disk with a resonant ring structure.  The spikes in the collision rate vs semi-major axis and eccentricity show that the collision rate is enhanced for particles in resonance, and also adjacent to the first order MMRs.  \label{collision_rate_vs_ae_figure}}
\end{center}
\end{figure}

The top panel in Figure \ref{collision_rate_vs_ae_figure} shows the collision rate per particle as a function of semi-major axis in the region of the resonant ring, a kind of continuum with spikes.  The continuum of the collision rate is higher for semi-major axis values close to 1 AU than for larger semi-major axis values because the resonant ring enhances the local collision rate by a factor of $\sim 2$, as shown in Figure \ref{collision_rate_faceon_figure}.  The spikes in the collision rate correspond to MMRs: the 2:1, 5:3, 3:2, 7:5, 4:3, 9:7, 5:4, etc., from right to left.  Most of the first order resonances (p+1:p) show a split peak, with higher collision rates at the edges of the resonance than at the center.  The split peaks may be caused by collisions between the grains just outside of the MMRs with the grains in the MMRs.  


The dashed line in the top panel of Figure \ref{collision_rate_vs_ae_figure} shows the classically-calculated collision rate, $1/t_{\rm coll} = \alpha\tau / t_{\rm orbit}$.  We set $\alpha=2.6$ so that the collision rates were equal at a semi-major axis of 1.5 AU.  This value of $\alpha$ is less than the estimate $\alpha \approx 4\pi$ by \citet{wdt99}, and hence our collision rate is lower, likely because we simulate only a single grain size whereas \citet{wdt99} consider a distribution of grain sizes and include catastrophic collisions caused by projectile grains much smaller than the target grain.

Our algorithm shows how the classically-calculated collision rate fails for disks with planets, since this approximation neglects collisional enhancement of particles in resonance and does not accurately approximate the average collision rate in the resonant ring structure.  It also cannot correctly reproduce the vertical structure of the collision rate, like that shown in the bottom-left panel of Figure \ref{collision_rate_faceon_figure}.  The failure of this approximation is even more dramatic than these figures show because the particles spend most of their time in the MMRs.

The bottom panel of Figure \ref{collision_rate_vs_ae_figure} shows the collision rate per particle as a function of eccentricity.  The particles attain an eccentricity of no more than $e \approx 0.2$ when they are launched.  So any particles with $e > 0.2$ in this plot must have had their eccentricities increased by resonant pumping or close encounters with the planet.  These particles have higher collision rates because they are typically located in the resonant ring, a region of higher density and larger velocity dispersion.  The small peaks in the collision rate at $e = 0.27$, $e = 0.31$, and $e = 0.37$ correspond to the maximum eccentricities that particles can obtain in the 5:4, 4:3, and 3:2 exterior MMRs \citep{bfm94}, and also correspond to localized regions of higher density within the resonant ring structure.  The data become noisy for $e > 0.38$ because relatively few particles end up with such large eccentricities, except in close encounters with the planet.

\begin{figure}
\begin{center}
\includegraphics[width=6.5in]{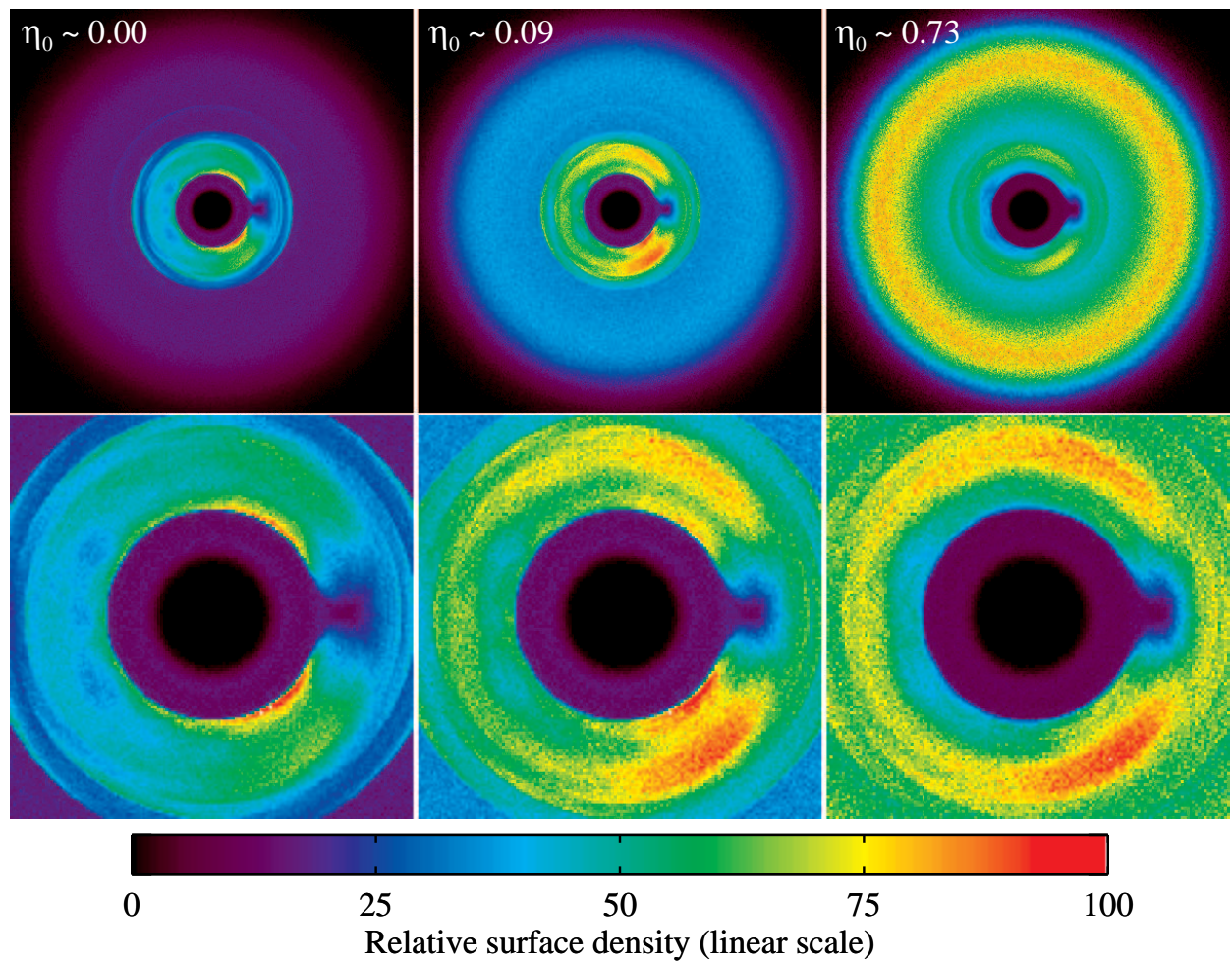}
\caption{Surface density as a function of $\eta_0$ for a disk with a ring structure caused by an Earth-mass planet at 1 AU viewed face-on.  The top row shows the entire disk, which extends out to 4.25 AU.  The bottom row shows zoomed-in views of the resonant ring structure.  Collisions reduce the sharp inner-edge feature of the resonant ring structure, smear out azimuthal structure, and de-emphasize the resonant ring while emphasizing the birth ring.  Even a low collision rate ($\eta_0 \ll 1$) can significantly alter resonant ring structures. \label{ring_vs_eta_figure}}
\end{center}
\end{figure}

Figure \ref{ring_vs_eta_figure} illustrates some of the morphological effects collisions can have on resonant ring structures.  The top row shows surface density histograms of a disk with a resonant ring structure for three different values of $\eta_0$, increasing from left to right.  The bottom row shows zoomed-in views of the resonant ring structures from each of the disks in the top row.  Each of the six histogram images was scaled independently to highlight differences in geometry.

Collisions de-emphasize the resonant ring and emphasize the birth ring.  They also change the morphology of the resonant ring.  In the collisionless case on the left, the resonant ring structure has a sharp inner edge with well-defined azimuthal and radial structure.  For the slightly collisional system ($\eta_0 \sim 0.09$), shown in the middle, the density at the resonant ring's inner edge is significantly reduced relative to the rest of the ring structure.  As we further increase the overall collision rate in the disk, the density of the inner edge of the resonant ring continues to drop relative to the rest of the ring, while azimuthal structures in the ring become smeared out.

Collisions reduce the density of the inner edge of the resonant ring for two reasons.  First, as previously shown, collision rates are higher in regions of higher density.  Second, grains that contribute to the inner edge of the resonant ring are typically older, having had their eccentricities pumped up while in resonance.  These older grains have had more time to collide with other grains.

\section{Particle fragmentation \label{particlefragmentationsection}}

Our algorithm, as described above, can handle seed models with multiple grain sizes with no modifications.  With a small modification the algorithm can also model fragmentation, i.e., collisions that produce daughter particles.  Here we describe a method for including particle fragmentation and present some preliminary results.

When two particles collide and produce fragments, the fragments are launched into new orbits such that the fragment velocity vectors are distributed around the center of momentum velocity vector of the colliding particles \citep[e.g.][]{kss05}.  Integrating the orbits of all of these fragments would be computationally prohibitive, so we make a numerical approximation: we limit the trajectories of the fragments we model to the recorded trajectories that already exist in the seed model.  To make this approximation work, our seed model must include a sufficiently wide range of initial conditions to ensure that trajectories are available that closely match the desired distribution of fragment trajectories.

To include particle fragmentation in our algorithm, we first run a seed model with several discrete particle sizes, each of which represents a range of sizes, and initially populate each size bin according to a mass distribution function \citep[e.g.][]{d69}.  Then we apply our collision algorithm with the following additional subroutine, implemented during the calculation of the collision depth (Equation \ref{collisiondepthequation}) for every record of each stream:
\begin{enumerate}
	\item Calculate the $k^{\rm th}$ record's contribution to the collision depth of the $i^{\rm th}$ record (c.f. Equation \ref{collisiondepthequation}), call it 
		\begin{equation}
			\tau_{{\rm coll,}i,k} = n_k \sigma_k | {\bf v}_i - {\bf v}_k | t_{\rm record}.
		\end{equation}
	Remember that the index $i$ refers to records in a stream while the index $k$ refers to records in a bin.
	\item Calculate the mass of particles removed from the $i^{\rm th}$ record by collisions with the $k^{\rm th}$ record, given by
		\begin{equation}
			\Delta M_{i,k} = m_i N_{i-1} \left(1-e^{-\tau_{{\rm coll,}i,k}}\right),
		\end{equation}
		where $m_i$ is the mass of a single particle in the $i^{\rm th}$ record.
	\item Record the center of momentum velocity vector of the colliding particles, given by 
		\begin{equation}
			{\bf v}_{{\rm COM},i,k} = \frac{m_i {\bf v}_i + m_k {\bf v}_k}{m_i + m_k}.
		\end{equation}
		We assume the difference between the fragment velocities and ${\bf v}_{{\rm COM},i,k}$ is small and use ${\bf v}_{{\rm COM},i,k}$ as the desired fragment velocity.
	\item Distribute the fragments by size according to a crushing law.
	\item Search within the local spatial bin for streams that closely match the grain size, $s$, and center of momentum velocity vector, ${\bf v}_{{\rm COM},i,k}$, of the fragments.  Once the appropriate streams are found, increase the numbers of particles in those streams to account for the fragments.
\end{enumerate}

The subroutine described above must be executed for every pair of records in every bin.  In practice, searching for an appropriate stream in which to put fragments during every interaction is computationally expensive; the algorithm run time scales as $\langle n_{\rm b}\rangle ^3$ instead of $\langle n_{\rm b}\rangle ^2$.  One could imagine many possible approximations that would reduce the amount of computer time spent searching for fragment streams.  For our preliminary models, we chose to place all of the fragments from the $i^{\rm th}$ record into a single mean center of momentum stream, with velocity
	\begin{equation}
		\langle {\bf v}_{\rm COM,i} \rangle = \frac{\displaystyle\sum_k \Delta M_{i,k} {\bf v}_{{\rm COM},i,k}}{\displaystyle\sum_k \Delta M_{i,k}}.
	\end{equation}
We leave a detailed investigation into the accuracy and efficiency of this approximation for future work.

We have implemented this particle fragmentation subroutine in our code and produced a simple model of a fragmenting disk with a resonant ring structure to illustrate the procedure at work.  For our seed model, we integrated the orbits of 2,500 120 $\mu\rm{m}$ grains ($\beta = 0.0023$) and 2,500 12 $\mu\rm{m}$ grains ($\beta=0.023$) and recorded the particle positions and velocities every 426 and 42 years, respectively, as they orbited a Sun-like star (sw$=0.35$) in the presence of an Earth-mass planet on a circular orbit at 1 AU.  We launched the grains from parent bodies with initial conditions identical to the second seed model described in Section \ref{seedmodelsection}.  For the purposes of illustration, we initially populated only the 120 $\mu\rm{m}$ grain streams and implemented a simple fragmentation scenario in which 120 $\mu\rm{m}$ grains shatter completely into 12 $\mu\rm{m}$ grains, conserving mass.  Fragments from 12 $\mu\rm{m}$ grains were deleted from the model.  We assumed all colliding grains were shattered completely, regardless of the mass or velocity of the target or projectile.  We scaled the number of 120 $\mu\rm{m}$ particles per stream such that $\eta_0 \sim 0.04$.

Figure \ref{fragmentationfig} shows the results of our fragmenting disk model.  The inset false-color image shows the face-on surface density of the disk.  The 120 $\mu\rm{m}$ (large) grains, shown in red, dominate the surface density exterior to the resonant ring structure.  The 12 $\mu\rm{m}$ (small) grains, shown in blue, dominate the surface density near the center of the disk.

The plot in Figure \ref{fragmentationfig} shows the disk mass distribution as a function of semi-major axis for each of the grain sizes in our simple model.  At the outer edge of the disk, near the birth ring of large grains at $\sim$4 AU, the large grains dominate the mass of the disk.  As the large grains spiral inward, collisional fragmentation reduces the disk mass in large grains and transfers that mass to the small grains.

Near 1.5 AU, the resonant ring structure enhances the collision rates of the large grains and therefore also the disk mass in small grains.  The spikes in the mass distribution function near 1 AU show that the two grain sizes populate different sets of MMRs.  The combined effects of collisional fragmentation and PR drag would cause the resonant ring structure, and the disk as a whole, to look different at different wavelengths, because grains of different sizes emit differently.  Figure \ref{fragmentationfig} shows that PR drag and resonant interactions can sort collisionally fragmenting grains by size, allowing smaller grains to spiral in to smaller circumstellar distances.

The radial distribution of dust grains produced by our model is analogous to that observed in the Fomalhaut disk.  \citet{s04} resolved the Fomalhaut disk at 24, 70, and 160 $\mu\rm{m}$ using the Multiband Imaging Photometer for \emph{Spitzer} (MIPS) and obtained a 17.5--34 $\mu\rm{m}$ spectrum with the Infrared Spectrograph (IRS).  Both the IRS spectrum and the 24 $\mu\rm{m}$ MIPS image reveal a compact source of dust likely interior to 20 AU, well inside of Fomalhaut's outer ring structure near 140 AU.  This compact source of dust, responsible for $\sim$0.7 Jy of flux at 24 $\mu\rm{m}$, is not seen in the 70 or 160 $\mu\rm{m}$ MIPS images, suggesting that the warm dust grains are inefficient at radiating at these wavelengths; the compact warm dust grains may be smaller in size than the grains near the outer ring.  Our preliminary fragmentation model shown in Figure \ref{fragmentationfig} suggests that collisional fragmentation of large grains triggered by MMRs may be the source of small dust grains.

\section{Summary}

We have developed a new algorithm to self-consistently treat collisions and resonant gravitational dynamics in dusty disks.  Our algorithm handles disks with multiple grain sizes and can be adapted to model particle fragmentation.  The algorithm uses the density and velocity distributions of a collisionless disk simulation to iteratively solve for the density distribution of a steady-state collisional disk.  The algorithm is applied after the simulation of the collisionless system, removing the need to re-integrate the equations of motion for disks with different collision rates, and can run on a single processor in $\sim 1$ hour.

We performed several tests to show that our algorithm arrives at a unique and correct solution for collisional disks with and without a resonant ring structure.  We showed that collisions can reduce the contrast of resonant ring structures, especially at the inner edge of the ring structure, and smear out azimuthal asymmetries.  We also showed that particle fragmentation triggered by resonant interactions can radially sort particles by size, producing smaller particles at smaller circumstellar distances.  This process may explain the population of warm dust found interior to Fomalhaut's ring \citep{s04}.

\begin{figure}
\begin{center}
\includegraphics[width=6.5in]{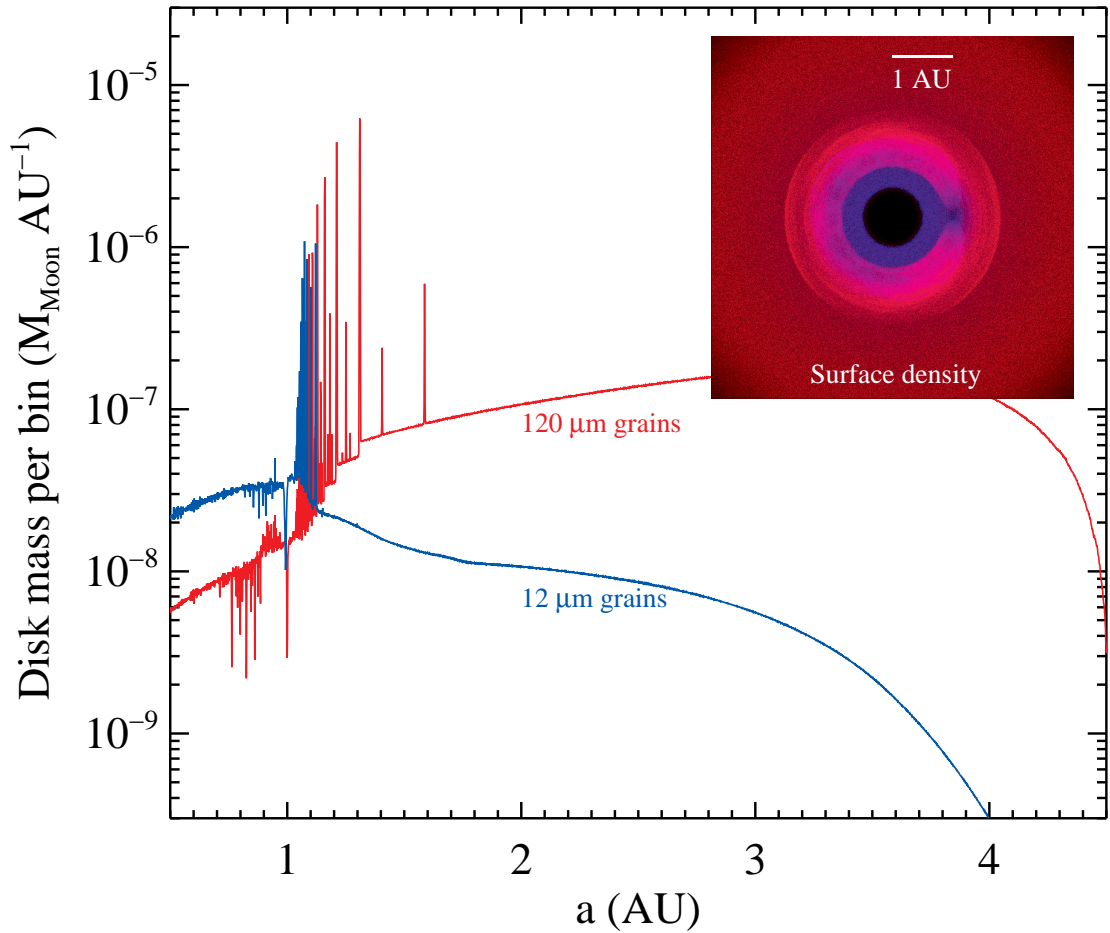}
\caption{Disk mass (in Lunar masses) as a function of semi-major axis for a disk of fragmenting grains in the presence of an Earth-mass planet at 1 AU orbiting the Sun.  The inset false-color image shows the face-on surface density of the disk.  MMRs near 1 AU trigger fragmentation, a process which may explain the population of small warm dust interior to Fomalhaut's resolved ring structure \citep{s04}. \label{fragmentationfig}}
\end{center}
\end{figure}

Our collisional grooming algorithm should allow us to accurately model and synthesize multi-wavelength images of observed debris disks, like Fomalhaut, Vega, and HR 4796A.  The algorithm enables us to investigate the effects that collisions have on dust disk morphology, such as asymmetries from clumping of parent bodies, resonant trapping of dust grains, and the radial sorting of grain sizes illustrated in Figure \ref{fragmentationfig}.  It should be useful for modeling long-lived grains in the solar zodiacal cloud and it should help us predict the morphology of ring structures in disks yet to be observed.

\acknowledgments

We thank the National Aeronautics and Space Administration (NASA) Goddard Space Flight Center's Graduate Student Researchers Program for funding this research and the NASA High-End Computing Program for granting us time on the Discover cluster.  We thank the International Space Science Institute in Bern, Switzerland, for support of this research.  Marc Kuchner also thanks the NASA Astrobiology program for support.

\newpage
\clearpage

\newpage
\clearpage

\newpage
\clearpage

\newpage
\clearpage


\newpage
\clearpage


\newpage
\clearpage
\begin{deluxetable}{cccc}
\tablewidth{0pt}
\footnotesize
\tablecaption{Benchmark Tests \label{benchmark_table}}
\tablehead{
\colhead{Number of Particles} & \colhead{Bin Size} & \colhead{Run Time per Iteration\tablenotemark{\dag}} \\ 
 & \colhead{(AU)} & \colhead{(min)} \\
}
\startdata
1,250 & 0.05 & 0.3 \\
2,500 & 0.05 & 0.8 \\
3,750 & 0.05 & 1.6 \\
5,000 & 0.05 & 2.7 \\
1,250 & 0.1 & 1.1 \\
2,500 & 0.1 & 4.2 \\
3,750 & 0.1 & 9.4 \\
5,000 & 0.1 & 16.7 \\
\enddata
\vspace{-0.1in}
\tablenotetext{\dag}{For a single 2.2 GHz CPU}
\end{deluxetable}

\newpage
\clearpage

\newpage
\clearpage

\newpage
\clearpage

\newpage
\clearpage



\end{document}